\documentclass[aas_macros,usenatbib,useAMS]{mn2e}

\usepackage[authoryear]{natbib}
\bibpunct{(}{)}{;}{a}{}{,}
\usepackage{graphicx}
\usepackage{color}
\usepackage{amssymb}
\usepackage{ulem}
\usepackage{natbib}
\usepackage{psfrag}

\date{}

\title[Interstellar scintillation is an indicator of astrometric stability]{Scintillation is an indicator of astrometric stability}
\author[R. G. Schaap et al.]{R. G. Schaap$^{1}$, S. S. Shabala$^{1}$\thanks{e-mail:Stanislav.Shabala@utas.edu.au}, S. P. Ellingsen$^{1}$, O. A. Titov$^{2}$, J. E. J. Lovell$^{1}$\\
$^{1}$School of Mathematics and Physics, Private Bag 37, University of Tasmania, Hobart, TAS 7001, Australia \\
$^{2}$Geoscience Australia, PO Box 378, Canberra, ACT 2601, Australia}

\begin{document}


\pagerange{\pageref{firstpage}--\pageref{lastpage}} \pubyear{2013}

\maketitle

\label{firstpage}

\begin{abstract}

We examine the relationship between astrometric stability and astrophysical properties in flat-spectrum radio-loud quasars making up the celestial reference frame. We use position determinations from geodetic VLBI measurements, and develop a new metric for source position stability. We then compare this quantity to two measures of source compactness: structure index, which probes structure on milliarcsecond scales; and the presence of interstellar scintillation associated with the quasar, which probes scales of tens of $\mu$as.

We find that persistent scintillators have greater position stability than episodic scintillators, which are in turn more stable than non-scintillators. Scintillators are also more likely to be compact on milliarcsecond scales, as measured by the structure index. Persistent scintillators are therefore excellent candidates for inclusion in any future realisation of the celestial reference frame. A list of these sources is presented in Appendix~\ref{app:persistentScints}.

We find that slow (characteristic timescale $>3$~days) scintillators have more stable positions than rapid scintillators, suggesting they are more compact. High-cadence, long term monitoring is therefore required to identify other members of this population of compact, high brightness temperature quasars.

\end{abstract}


\section{Introduction}
\label{sec:intro}

The central regions of all massive galaxies are believed to contain supermassive black holes \citep{KormendyRichstone95,MagorrianEA98,KormendyEA09}.  When large amounts of material are being accreted onto these central black holes, the conversion of gravitational potential energy into radiation, either from an accretion disk or through launching of relativistic jets \citep{PenroseFloyd71,BlandfordZnajek77,BlandfordPayne82}, produces more energy than the combined output of all the stars in the host galaxy \citep[e.g.][]{WatsonEA08}. The most extreme examples of these active galactic nuclei (AGN) are quasars; these are the most luminous individual objects in the Universe.  The high luminosity of quasars means that they are interesting to study both for intrinsic reasons -- they give insights into physics in extreme conditions which cannot be reproduced in a laboratory -- and they can also be used as a tool in other astrophysical investigations. The high luminosity of quasars means that they are visible at extremely large distances, indeed there are no quasars in the local Universe and they are predominantly observed at $z > 1$, when galaxy interactions \citep[thought to be responsible for triggering much of the quasar activity,][]{HopkinsEA06,ShabalaEA12} were more common.

Because they are found at very large distances, quasars provide a good approximation to an isotropic distribution of bright, fixed point sources. For this reason, they are used to define the International Celestial Reference Frame \citep[ICRF;][]{MaEA98}. The highest angular resolution astronomical observations that can be undertaken use the technique of Very Long Baseline Interferometry (VLBI), at radio wavelengths.  When accurate calibration observations are undertaken and applied it is possible to measure the relative position of a foreground object with respect to a reference source (a distant quasar), to an accuracy of approximately 10~$\mu$as.  Astrometry with this level of accuracy has been used to measure the trigonometric parallax of sources at distances $>$~5~kpc \citep{ReidEA09}, and place limits on the speed of propagation of the gravitational field \citep{FomalontEA03}. The same ICRF can also be used to make measurements of the relative position of objects on Earth (geodesy). Most frequently observed ICRF quasars have position uncertainties of $20-50$~$\mu$as, and the reference frame itself is stable at the level of 10~$\mu$as \citep{MaEA09}. Geodetic VLBI observations provide the only means of measuring simultaneously terrestrial parameters such as the Earth's orientation and length of day \citep{MaEA98,MaEA09}.

The accuracy of specific astrometric or geodetic measurements depends on a range of factors \citep[see e.g.][]{ShabalaEA13}, but one of these is the accuracy with which the positions of the reference sources have been determined and the degree to which they deviate from an ideal point source.  At radio frequencies the emission from distant quasars arises from highly relativistic jets of material which lie at angles close to our line of sight.  In some sources the radio emission is observed to evolve on milliarcsecond scales on timescales of months to years \citep[e.g.][]{ListerEA09,GrossbergerEA12}.  These sources clearly deviate from an ideal point source and hence when attempting high precision astrometric or geodetic observations it is very important to identify which reference sources (quasars) are best suited for this task.

The International VLBI Service \citep[IVS;][]{SchuhBehrend12} undertakes regular VLBI observations to measure a range of geodetic parameters and the current implementation of the ICRF, known as the ICRF2 \citep{MaEA09}, is based on these observations.  In order to improve the accuracy of the geodetic measurements the IVS continually looks to increase the density, uniformity and quality of the quasars used in the ICRF determination.  In particular, the small number of radio telescopes in the southern hemisphere compared to the northern hemisphere means that the radio properties of southern quasars are generally less well determined than those in the north and far fewer suitable sources have been identified, particularly south of -40$^\circ$ \citep[e.g.][]{OjhaEA04,FeyEA04,FeyEA06}.  VLBI observations typically have an angular resolution of around 1 milliarcsecond, with the highest angular resolution observations (around 100~$\mu$as) being obtained from Earth-based observations at frequencies of 22~GHz and above, or arrays including a space-based antenna.  The highest angular resolution information on AGN is obtained indirectly through observations of the interstellar scintillation (ISS) which occurs for the most compact sources \citep[e.g.][]{DennettThorpedeBruyn00,LovellEA03,CarterEA09}.  For an AGN to exhibit ISS, a substantial fraction of its intensity must arise on angular scales comparable to the angular scale of the scattering screen (as viewed by the observer).  Most AGN do not show ISS at centimetre radio wavelengths, however a significant fraction of flat-spectrum AGN do, indicating that a significant fraction of their emission arises on angular scales of less than 10--50~$\mu$as \citep{LovellEA08}.  This is supported by VLBI observations which show that scintillating sources are more core-dominated than non-scintillating AGN \citep{OjhaEA04,OjhaEA06}.  This suggests that AGN which are observed to show ISS are promising candidates for future expansion of the sources used to define the ICRF.

Currently geodetic VLBI analysis primarily measures the quality of individual sources through a quantity known as the structure index (see Section~\ref{sec:SI_data}), which is determined from radio images of the quasars.  For many of the quasars there are a limited number of images available (often only one), and the images typically have angular resolutions of greater than 1~milliarcsecond.  So there is the potential for the structure index to poorly reflect how compact (or otherwise) the source is, if the emission at the epoch the images were made is atypically compact (or extended) for that particular source.  It's also unclear to what degree the milliarcsecond scale emission reflects whether or not the source is compact on angular scales 10 to 100 times smaller.  

The IVS observations now span a time baseline of 30 years, and for each experiment, the output products from the geodetic analysis include the position of each of the sources observed.  Many sources have been observed in hundreds of independent experiments, hence the quality of the source for geodetic, or astrometric purposes can be directly assessed from the repeatability of the position determination.  In order to improve the selection of sources for future geodetic and astrometric VLBI observations, and to find new candidates reference sources, we have undertaken a comparison of three measures of quasar compactness: {\it (i)} time series of source astrometric positions as estimated from geodetic VLBI observations; {\it (ii)} structure indices derived from VLBI images of these sources; and {\it (iii)} their scintillation properties.

\section{The data}
\label{sec:data}

\subsection{Geodetic positions}
\label{sec:geo_pos_data}

Global IVS observations were used to construct the quasar position time series. The data were processed using the OCCAM geodetic analysis software package \citep{TitovEA04}, which adopts the least squares collocation technique to calculate Earth Orientation Parameters, and antenna and source positions for each 24-hour observing session. The analysis was carried out according to the IERS 2010 Conventions \citep{PetitLuzum10}, with the exception of tidal atmospheric loading which was not implemented. In total, 145,636 positions of 3054 sources were estimated in 3804 geodetic sessions over a 26-year period from 1984--2009. We note that 96 percent of the observations have taken place since 1990, and excluding the (slightly lower quality) pre-1990 data does not affect our results.


\subsection{Structure indices}
\label{sec:SI_data}

Most quasars are not point-like, and the quasar brightness distribution therefore contributes an additional term to the interferometer visibility phase as measured on any given baseline. The sign and magnitude of this phase term depend on the projected quasar structure as viewed by the baseline. In geodetic VLBI, the fundamental observable is the so-called group delay, defined as the slope of phase across a 750 MHz bandpass between 8.2 and 8.95~GHz \citep[e.g.][]{SchuhBehrend12}. \citet{Charlot90} developed the formalism for estimating the effects of quasar structure on geodetic measurements, by calculating the change in the group delay as a function of quasar brightness distribution and the baseline vector. \citet{FeyCharlot97,FeyCharlot00} defined the {\it structure index} as the logarithm of the median time delay due to quasar structure observed with all possible Earth-bound baselines, ${\rm SI} = 1 + 2 \log (\tau / {\rm ps})$. Sources with ${\rm SI} > 3$ are considered unsuitable for geodetic VLBI \citep{MaEA09}. By definition, the structure index is an averaged quantity which does not include any orientation information: the structure contribution to the group delay will be greatest when the source elongation direction is parallel to the observing baseline, and smallest (in fact, zero) when these two vectors are orthogonal.

VLBI images exist for over 4000 sources \citep{FeyCharlot97,FeyCharlot00,PetrovDatabase}, many of these in multiple epochs. This is crucial, as quasar structure is known to vary significantly on timescales of months, with quasi-regular outbursts and component proper motions of hundreds of $\mu$as per year \citep{ListerEA09}. In the present work we use the 701 median structure indices for ICRF2 quasars calculated by the Bordeaux group\footnote{http://www.obs.u-bordeaux1.fr/BVID} and presented in \citet{MaEA09}. 

\subsection{Scintillation}
\label{sec:scintillation_data}

The Micro-Arcsecond Scintillation-Induced Variability (MASIV) survey \citep{LovellEA03,LovellEA08} used the Very Large Array (VLA) to study the flux density variability of 443 flat spectrum ($\alpha>-0.3$) quasars at 4.9 GHz. All the quasars were compact (at least 95 percent of their 8.5~GHz VLA flux density in an unresolved sub-arcsecond component), and brighter than 100~mJy. The sources were monitored in four epochs, each three or four days in duration, evenly spaced over the course of one year. On average, flux density measurements of each source were made every two hours when the source was observable. In each epoch, sources were identified as either scintillators or non-scintillators based on whether they displayed significant changes in total flux density.

Because MASIV sources are selected to be compact on arcsecond scales, many of these sources are ICRF2 quasars. We can therefore quantify the scintillation properties of many ICRF2 sources for which we also have position and/or structure index measurements. We discuss the cross-matching of our three samples in Section~\ref{sec:results}.

\section{Position stability}
\label{sec:analysis}

\subsection{Source positions from geodetic observations}
\label{sec:geo_pos_explanation}

\begin{figure}
\includegraphics[width=84mm]{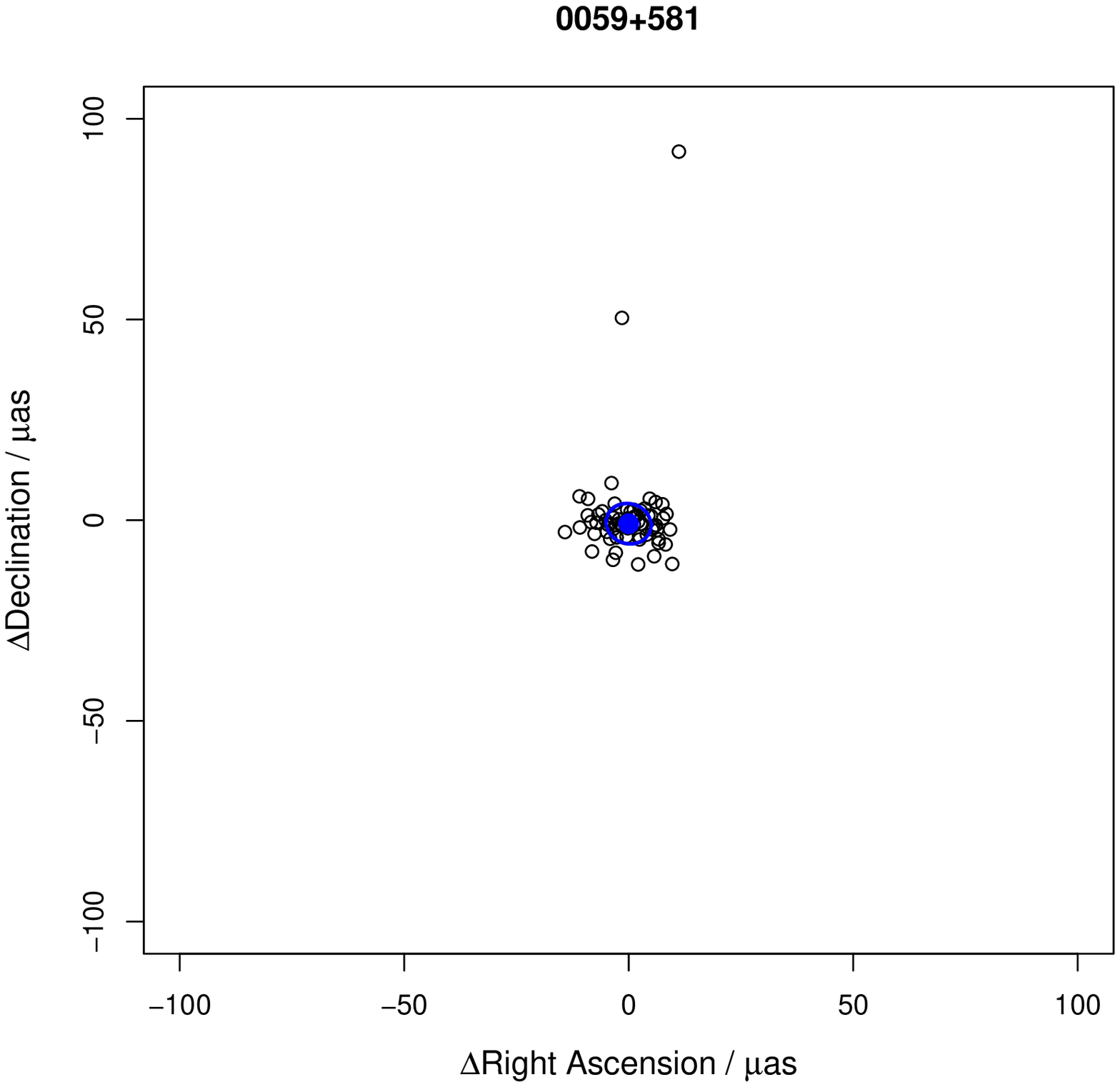}
\includegraphics[width=84mm]{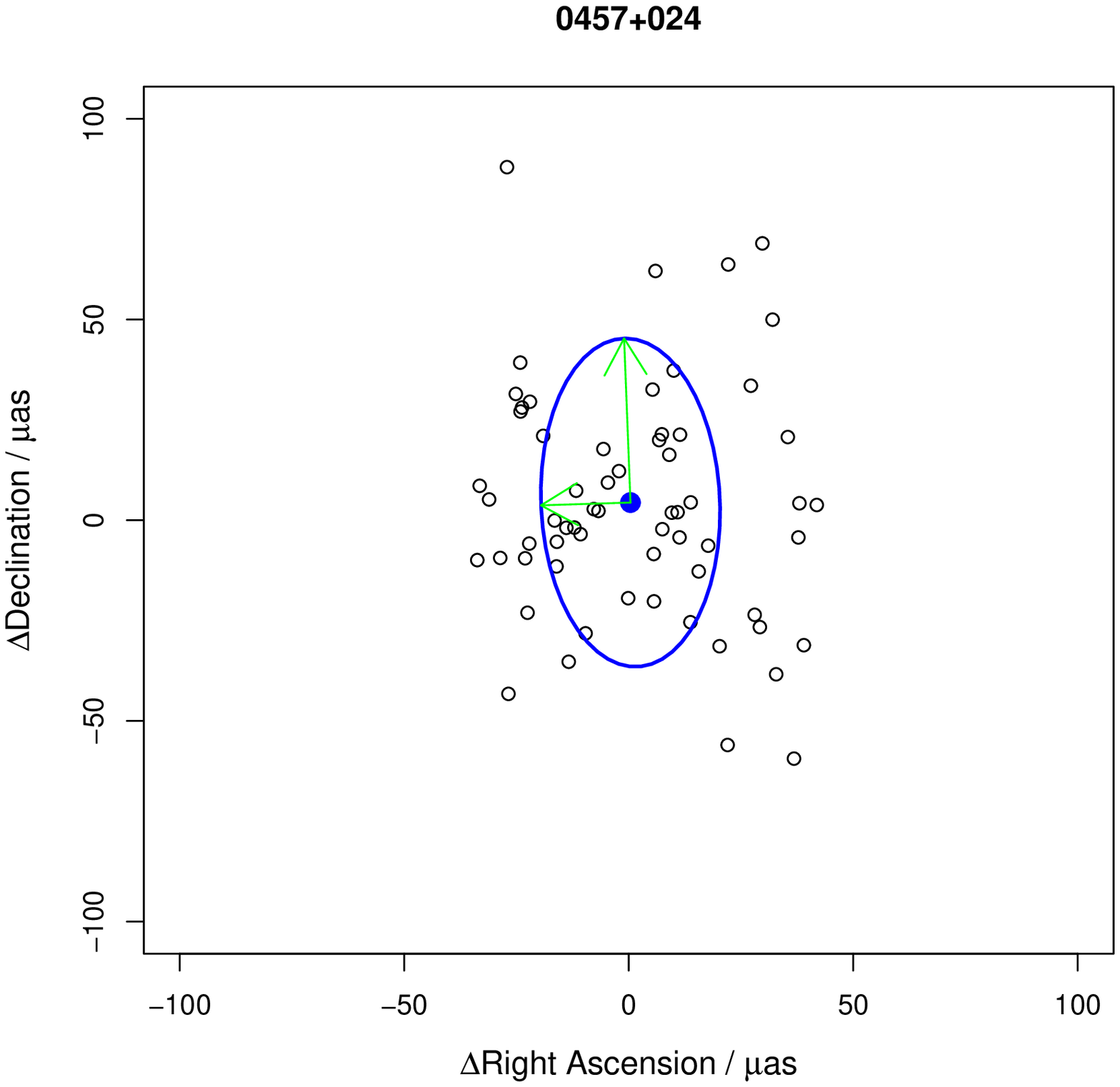}
\caption{Scatter in source position about the mean coordinates for two ICRF2 quasars. {\it Top panel}: $0059+581$, a persistent scintillator; {\it bottom panel}: $0457+024$, a non-scintillator. Each point denotes a measurement derived from a 24-hour geodetic session. The 68 percent confidence ellipses are shown with major and minor axes. To illustrate the difference in position scatter, we show the same number of points (69) for the two sources. The 69 points shown for $0059+581$ were randomly selected from the full time series of 1964 observations. The confidence ellipses are derived from all available data.}
\label{fig:pos_scatter}
\end{figure}

For a point-like quasar observed in the absence of any source of astrometric error, the position measurement derived using OCCAM would coincide exactly with the catalogued position for each individual observation. In reality, source coordinates rarely agree with either the catalogued position or between sessions. This position scatter, shown in Figure~\ref{fig:pos_scatter}, comes about due to three factors. Instrumental errors such as clock variability, both systematic and stochastic, are partially absorbed in source position estimates. A second factor is imperfect calibration of atmospheric and geophysical effects \citep[e.g. non-tidal ocean loading;][]{BoehmEA09}. For a sufficiently large number of uniformly scheduled observations, the scatter due to instrumental and atmospheric effects should be the same for all sources. Therefore, most relevant for the present work is the effect of source structure on geodetic position measurements. The estimated position of an extended radio source will vary depending on the orientation of source structure and observing baseline. This means that even without other sources of error, by virtue of observing the same non-point-like quasar with different VLBI networks and schedules, different positions will be estimated. This allows us to use source position scatter (from now on referred to as position stability) as a measure of how point-like a radio source is, or alternatively, how much structure it has. 

\begin{figure}
\includegraphics[width=84mm]{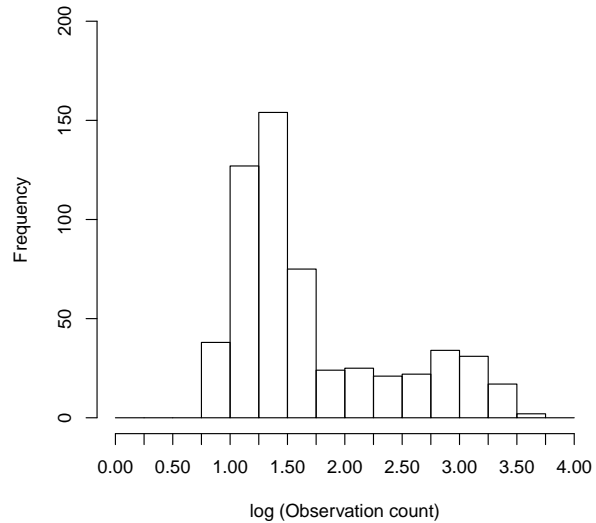}
\caption{Number of geodetic 24-hour sessions for each source.} 
\label{fig:counts_dist}
\end{figure}

In order to obtain meaningful position stability results, only the 570 ICRF2 sources with at least eight position measurements were included in our analysis. We note that we obtain qualitatively similar results (but at lower statistical significance) if the cut is made at 30 observations. The distribution of observation counts is shown in Figure~\ref{fig:counts_dist}. The majority of sources were observed between 10 and 30 times.

\subsection{Classification of positional stability variables}
\label{sec:classify_pos}

In order to define a quantitative measure of source position stability we fitted confidence ellipses to source positions using the statistical program {\texttt{R}}. An example is shown in Figure 1. Here the major and minor axes of the ellipse correspond to a 68 percent confidence interval. The shapes and orientations of each of these confidence ellipses were obtained through a 2x2 covariance matrix of source position data. This covariance matrix was evaluated under the assumption that the data came from a multivariate {\it t}-distribution; this method proved most robust to outliers. We note that all observations were used in our analysis, regardless of how far the measured position was from the catalogued source position. Although OCCAM provides formal uncertainties on individual position measurements, these are often $>$~100~$\mu$as \citep[see also Section 14 of][]{MaEA09}; we therefore did not use this information. We further note that \citet{MaEA09} argue that the scatter in source position time series dominates over formal uncertainties for most sources.

For each source, the major and minor axis lengths correspond to the square root of the eigenvalues of the covariance matrix. The directions of both these axes are given by the corresponding eigenvectors. The major and minor axes for two ICRF2 sources are displayed as green arrows in Figure~\ref{fig:pos_scatter}. The orientation of the confidence ellipse is determined by the position angle, defined as the angle of the major axis vector from zero in right ascension.

\section{Results and Discussion}
\label{sec:results}

\subsection{Position stability indicators}
\label{sec:pos_inst}

We determined the elongation ratio for each position ellipse, defined as the ratio of major and minor axis lengths. The majority of sources were found to have elongation ratios below three, although a third of the sources were more extended in one direction than this value. 

\begin{figure}
\includegraphics[width=84mm]{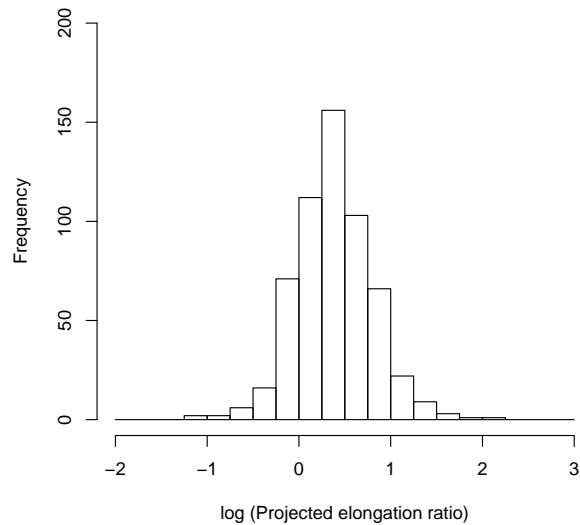}
\caption{Distribution of projected elongation ratios. Position scatter is significantly higher in Dec than RA.} 
\label{fig:proj_elongation_ratios}
\end{figure}

To test for possible biases in source position estimation, we calculated the {\it projected} elongation ratio for all sources. This was defined as the ratio of the projections of the confidence ellipse onto the declination and right ascension axes.  This quantity gives a measure of which coordinate a source is less stable in: for example, a source with a projected elongation ratio of three has a position which is three times more variable in declination than in right ascension. The distribution of projected elongation ratios is shown in Figure~\ref{fig:proj_elongation_ratios}. Clearly, the majority of our sources are more variable in declination rather than in right ascension, with the median projected elongation ratio being 2.3. This is expected due to the typical IVS network geometry producing synthesised beams that are more elongated in declination. 

This declination bias also affects our measurements of position angle. We found most sources to have position angles close to 90$^\circ$, corresponding to the confidence ellipse aligning with the declination axis; or  0$^\circ$, corresponding to the right ascension axis. This preference for elongation in one coordinate is an artefact of the geodetic solution: like all geodetic analysis packages, OCCAM estimates source positions in a least-squares sense, and the right ascension and declination coordinates are treated as separate parameters. Therefore, any position shift due to source structure is often absorbed in one coordinate but not the other.

In view of these results, both absolute and relative orientations of the major and minor position axes were deemed unreliable. In what follows, we only use the major axis length as a measure of source position stability.

\subsection{Reference frame in the southern hemisphere}
\label{sec:ICRF_south}

\begin{figure}
\includegraphics[width=84mm]{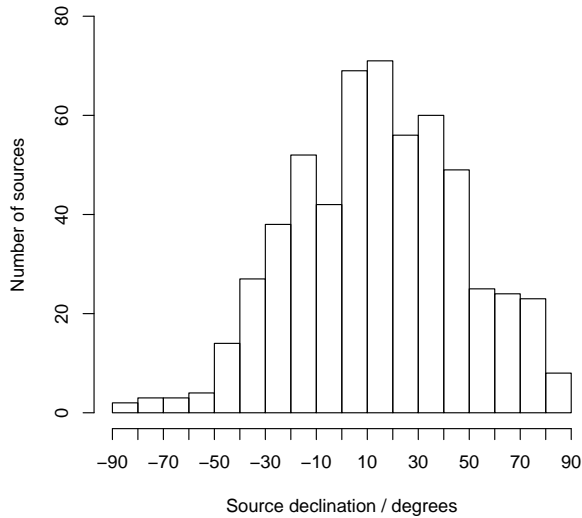}
\caption{Declination distribution of observed sources. There is a deficit of ICRF2 quasars in the southern hemisphere.} 
\label{fig:dec_dist}
\end{figure}

As discussed in Section~\ref{sec:intro}, a major limitation of the current realisation of the International Celestial Reference Frame, ICRF2, is its deficiency below a declination of $-40^\circ$. Figure~\ref{fig:dec_dist} shows that only 185 of 570 sources in our sample are located in the southern hemisphere, and only 26 sources at declinations below $-40^\circ$. At present, well-characterized geodetic sources are primarily ones that can be observed by the dense, long-established northern hemisphere network of VLBI antennas.

\begin{figure}
\includegraphics[width=84mm]{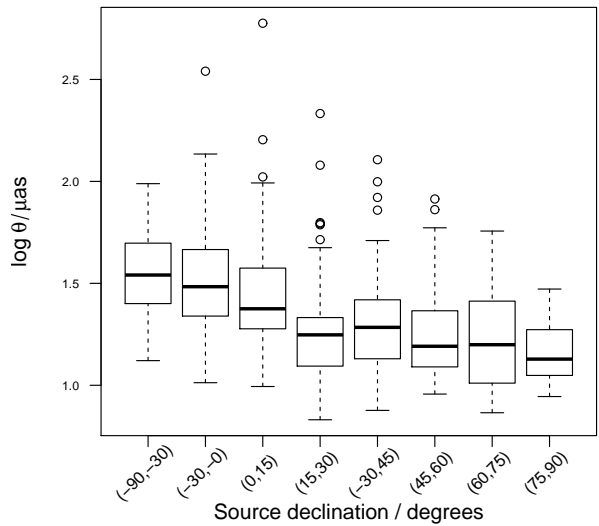}
\caption{Source position stability as a function of declination. The median major axis length of each declination group is displayed as a thick line, while 25th and 75th percentiles of the major axis distribution are shown with a box. Dashed lines indicate the 95 percent confidence interval. Major axis lengths are in microarcseconds, and thus a value of $\log \theta=1.0$ corresponds to 10~$\mu$as, and $\log \theta =2.0$ to 100~$\mu$as. There is an equal number of sources in each bin. There is a clear decrease in position stability at Dec~$<0$. There is no such trend for northern sources, apart from a marginal worsening of position stability at $0<$~Dec~$<15^\circ$; this is due to astrometry of equatorial sources being notoriously difficult.} 
\label{fig:maj_vs_dec}
\end{figure}

Figure~\ref{fig:maj_vs_dec} shows that southern ICRF2 sources have poorer position stability than their northern counterparts. A similar result has previously been reported by \citet{MaEA09}. Of course there is no physical reason for quasars at low declination to be less compact.  There would seem to be two factors contributing to poorer position stability for southern sources.  One is that a smaller number of compact quasars has been identified in the southern hemisphere studies to date (largely due to the smaller number of quasar studies in the south). The other factor has to do with observing network geometry: because the VLBI arrays in the south are more sparse, position determinations using these facilities are inherently less accurate.  This suggests that improving the determination of the ICRF in the south requires not only a densification of the number and quality of quasars, but also of the infrastructure used to make these measurements.  The recent commissioning of the AuScope geodetic VLBI array in Australia \citep{LovellEA13} and the Warkworth antenna in New Zealand \citep{WestonEA13} will address this issue and improve astrometry and geodesy in the southern hemisphere.

\subsection{Position stability and structure index}
\label{sec:pos_SI}

\begin{figure}
\includegraphics[width=84mm]{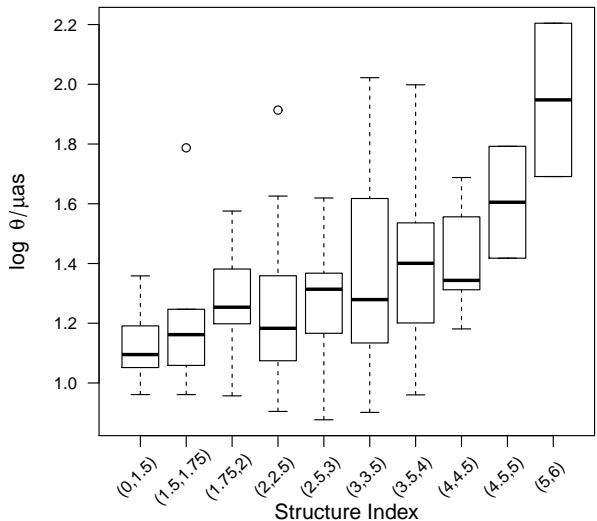}
\caption{Major axis length and structure index. High structure index sources have less stable positions, while the most compact sources are more positionally stable as expected. The relationship breaks down at intermediate structure indices (see text). Only five sources had structure indices $>4.5$, and hence the 95 percent confidence intervals are not shown for these objects.}
\label{fig:pos_vs_SI} 
\end{figure}

A well-established measure of source structure is the structure index. In Figure~\ref{fig:pos_vs_SI} we compare our major axis length stability indicator with structure index for the 168 sources which have both structure index and position stability measurements available. The two methods agree well, and are consistent with the results of \citet{MaEA09} who find that the median structure index correlates with a positional stability index $p$, which is calculated from a combination of formal errors and weighted root mean square (wrms) scatter in position. We note that this $p$ metric of \citet{MaEA09} is not reliable for a number of sources known to have strongly varying positions (see their Section~11.1.1) and we therefore do not use $p$ as a measure of position stability in the present work. 

Interestingly, Figure~\ref{fig:pos_vs_SI} shows that the correlation between the structure index and position stability is most obvious for very compact (SI~$<2$) and extended (SI~$>3.5$) sources. In other words, major axis stability appears to be good at picking out very stable and very extended sources, but not ones with some structure. We test this statistically by comparing compact (SI~$<1.5$), somewhat extended ($1.75 \leq$~SI~$<3.5$) and very extended (SI~$\geq 4$) sources. We find that these three populations are significantly different at the 95 percent level, as given by the Wilcoxon-Mann-Whitney test. On the other hand, splitting up the middle group into sources with structure indices $1.75-2.5$ and $2.5-3.5$, we find no statistically significant difference between these populations at the 95 percent level (p-value of 0.08). Structure indices are defined in terms of a {\it median} additional delay due to source structure \citep{FeyCharlot97} on all Earth-bound baselines. The averaging is done over all baselines, and all observing epochs. Many (perhaps most) flat-spectrum quasars are known to evolve on timescales of years \citep[e.g.][]{ListerEA09}, and will therefore exhibit jet components (i.e. structure) in some epochs but not others. The median structure index is therefore not expected to be a good indicator of positional stability for these sources. On the other hand, the very compact (SI~$<2$) sources are always core-dominated; while the very extended (SI~$>3.5$) sources always have a significant contribution to flux density from the jet component. The relative constancy of source structure in these objects accounts for the good correspondence between structure index and positional stability\footnote{Additionally, structure indices of 2 and 3 correspond to structure at the 20 and 60~$\mu$as level on a 10,000 km baseline, respectively. Astrometry at this level is very difficult with VLBI imaging, which can further reduce the reliability of structure indices in this range.}. 

\subsection{Position stability and scintillation}
\label{sec:pos_scint}

Both source position stability \citep[e.g.][and references therein]{MaEA09} and structure index \citep{FeyCharlot97,FeyCharlot00} have previously been used to quantify the suitability of a given quasar for geodesy. Astrophysically, the expectation is that scintillating radio-loud quasars should be compact (see Section~\ref{sec:intro}) and therefore suitable for geodesy. Furthermore, they should show low structure indices and stable source positions.

\subsubsection{Number of scintillating epochs}

\begin{figure}
\includegraphics[width=84mm]{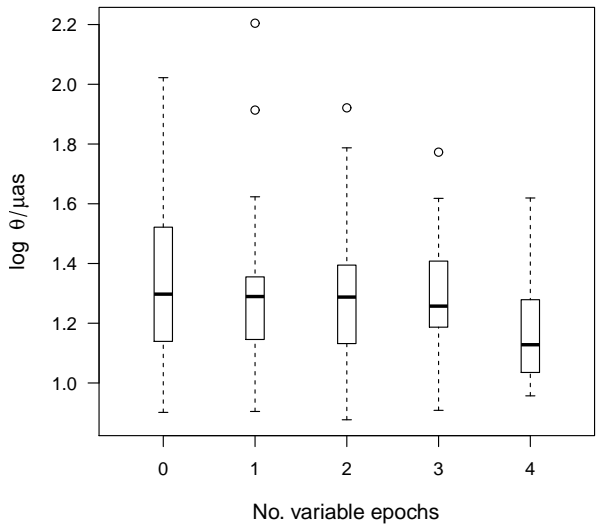}
\caption{Position stability and the number of scintillation epochs. Persistent scintillators show the greatest position stability, followed by episodic scintillators. Non-scintillators are least stable. Position stability indicators are as in Figure~\ref{fig:maj_vs_dec}.} 
\label{fig:pos_scintEpochs}
\end{figure}

Figure~\ref{fig:pos_scintEpochs} investigates the position stability of 170 MASIV quasars with eight or more position measurements. Here, sources are grouped by the number of epochs ($0 - 4$) in which they showed scintillation. In total, 66 sources did not scintillate at all;  39 sources scintillated in one epoch, 34 in two epochs; 17 in three epochs; and 14 sources were persistent scintillators (i.e. showed scintillation in all four epochs). Figure~\ref{fig:pos_scintEpochs} shows persistent scintillators (group 4) to have the smallest position major axes; in other words, these sources are the most stable. Conversely, flat-spectrum quasars that didn't scintillate in any epoch (group 0) are least positionally stable. Episodic scintillators (groups 1, 2 and 3) are between these extremes, and are statistically indistinguishable\footnote{This is not altogether surprising, since episodic scintillation is usually only just detected, and sources can often be misclassified as non-variable in other epochs due to the scintillation having a low modulation index.}. On the other hand, the difference between persistent, episodic, and non-scintillators is significant at the 95 percent level. Persistent scintillators clearly show more stable positions than both episodic scintillators, and non-scintillators.

\subsubsection{Modulation indices}

\begin{figure}
\includegraphics[width=84mm]{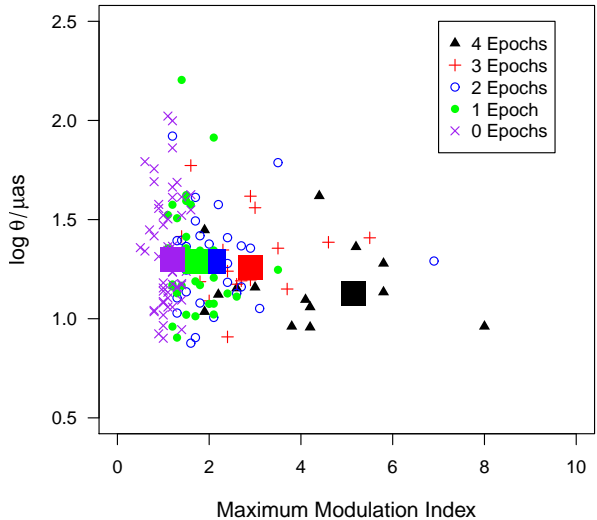}
\caption{Position stability and maximum modulation index (in percent). Different colours represent the number of epochs in which the source showed scintillation. Stars show median values for each bin. Persistent scintillators show the highest modulation indices, and are most stable. Non-scintillators have the lowest modulation indices, and are least stable. Modulation index is a good statistical estimator of source stability, and therefore compactness.} 
\label{fig:pos_vs_mi}
\end{figure}

A quantitative measure of source compactness is provided by the modulation index, defined as the ratio of rms to average flux density. Sources with high modulation indices are interpreted to have a higher fraction of flux density in a scintillating, compact component \citep[e.g.][]{LovellEA08}. Thus, compact sources are expected to not only scintillate at all epochs, but also to show higher modulation indices than less compact objects. Figure~\ref{fig:pos_vs_mi} considers the relationship between position stability and a source's maximum modulation index over the four epochs. We chose to consider the maximum, rather than mean, modulation index because this method does not artificially raise the modulation index of persistent scintillators in comparison with episodic ones. Persistent scintillators (black points) have both the highest maximum modulation indices and smallest major axis lengths. Conversely, the non-scintillators (purple) have the smallest maximum modulation indices and the largest major axis lengths. These findings are consistent with the results of Figure~\ref{fig:pos_scintEpochs}.

\subsubsection{Scintillation timescales}

\begin{figure}
\includegraphics[width=84mm]{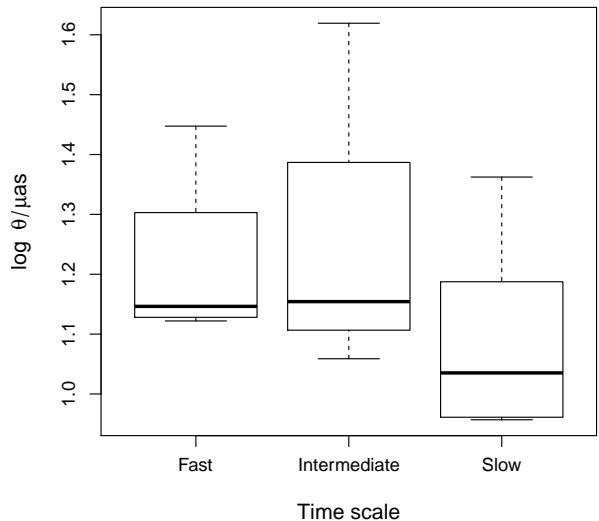}
\caption{Positional stability of persistent scintillators as a function of scintillation timescale. Slow scintillators have more stable positions than fast or intermediate scintillators. Symbol meaning is as in Figure~\ref{fig:maj_vs_dec}.} 
\label{fig:pos_tchar}
\end{figure}

Our final analysis of the link between scintillation and position stability concerns scintillation timescales. \citet{LovellEA08} classified scintillators into three groups: slow (characteristic timescale $t_{\rm char}>3$ days), intermediate ($0.5 < t_{\rm char} \leq 3$ days) and fast ($t_{\rm char} \leq 0.5$ days). Figure~\ref{fig:pos_tchar} shows that  slow persistent scintillators have significantly better position stability than fast or intermediate scintillators. This is because fast scintillation requires a relatively nearby scattering screen and these typically have high angular velocities \citep{DennettThorpedeBruyn00,BignallEA03}. Such screens are uncommon, but because they are unusually close the Fresnel scale can be quite large, causing even relatively extended components to scintillate. On the other hand, slow scintillation is associated with distant screens that have low angular velocities, which will only allow the most compact sources to scintillate \citep{TurnerEA12}.  Hence ISS with long characteristic timescales at centimetre wavelengths is one of the most reliable means of finding compact, high brightness temperature quasars.  These sources should, generally speaking, be very good reference sources for astrometric or geodetic purposes.

\subsection{Structure Indices and scintillation}

\begin{figure}
\includegraphics[width=84mm]{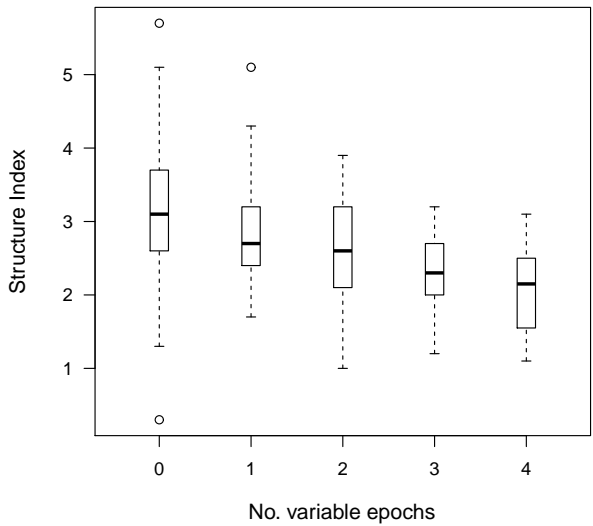}
\caption{Structure indices of MASIV scintillators, split by number of scintillation epochs. Frequent scintillators have lower structure indices. Symbol meaning is as in Figure~\ref{fig:maj_vs_dec}.} 
\label{fig:scint_SI}
\end{figure}

Having demonstrated that ISS is a good indicator of source position stability, in this section we investigate the relationship between scintillation and structure as measured by VLBI imaging. Previous investigations of a similar nature have been undertaken by \citet{OjhaEA04,OjhaEA06}, who used three different measures of compactness obtained from VLBA images of MASIV sources.  The three measures used by \citet{OjhaEA04} were the core fraction (the fraction of the flux density from the milliarcsecond scale images contained within a single synthesised beam), the flux-weighted radial extent and the unweighted radial extent.  They found that scintillating sources are significantly more compact than non-scintillating sources for each of these different measurements.  The structure index collated by \citet{MaEA09} provides a further alternative measure of source compactness. In total, 153 MASIV sources with at least eight position measurements had available structure index information.  Figure~\ref{fig:scint_SI} compares the structure indices of MASIV quasars binned according to the number of scintillation epochs. Persistent scintillators have the lowest structure indices, followed by episodic scintillators. Non-scintillators show the most structure in VLBI images. The differences between non-scintillators, 1 or 2 epoch scintillators, and 3 or 4 epoch scintillators, are statistically significant at the 95 percent level. This is qualitatively consistent with the results of \citet{OjhaEA04,OjhaEA06}. The advantage of our dataset compared to this earlier work is that we probe multiple scintillation epochs, allowing a separation of scintillators into persistent and episodic.

\begin{figure}
\includegraphics[width=84mm]{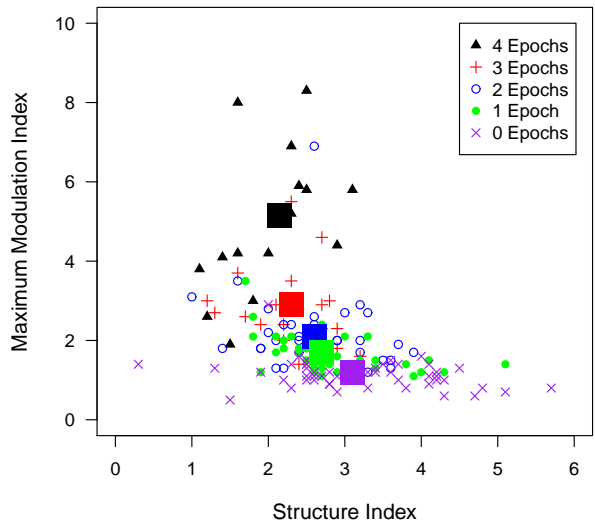}
\caption{Maximum modulation index correlates inversely with structure index. Symbol meaning is as in Figure~\ref{fig:pos_vs_mi}.}
\label{fig:mi_SI} 
\end{figure}

The relationship between the maximum modulation index and structure index is shown in Figure~\ref{fig:mi_SI}. Despite significant scatter\footnote{We re-iterate that for a source to scintillate, it must be both compact {\it and} have an appropriate scattering screen. Thus, many compact sources will not scintillate simply due to the absence of such a screen.}, medians show that highly variable sources have lower structure indices, in addition to having greater position stability (Figure~\ref{fig:pos_vs_mi}). We note that the number of scintillation epochs is correlated with both the structure index and maximum modulation index. In other words, frequent scintillators on average scintillate more strongly and are also more compact than rare (or non-) scintillators.

\begin{figure}
\includegraphics[width=84mm]{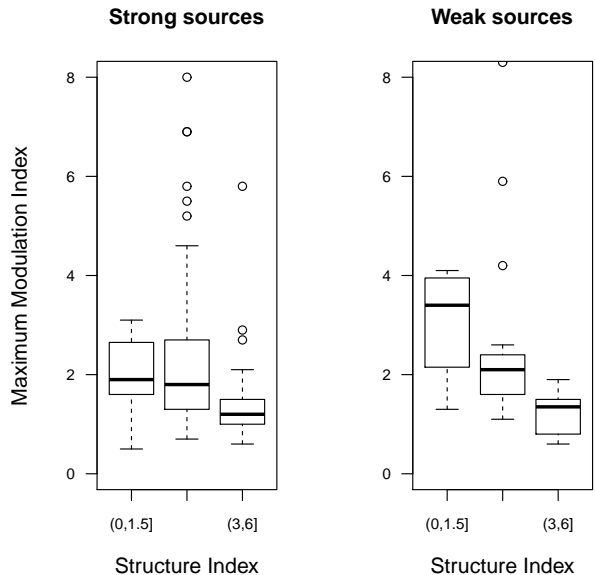}
\caption{Maximum modulation index -- structure index relation for weak ($<500$ mJy at 8.5 GHz) and strong scintillators.} 
\label{fig:mi_SI_weakStrong}
\end{figure}

\citet{LovellEA08} found that weaker sources were more likely to scintillate. They argue that the Inverse Compton limit on the source brightness temperature means that strong sources must be less compact than weak ones\footnote{There is also a slight selection effect: due to the constant noise floor, weak sources need to exhibit a higher level of variability in order to be classified as scintillators. However, this effect is not important for our flux density cut of 500~mJy \citep[see Figure~1 of][]{LovellEA08}. Even well above the noise limit, strong sources exhibit a deficit of high-amplitude ($>$ few percent) scintillators compared to weak sources.}. We have split up our sample of all sources that scintillated in at least one epoch into two subsets, with 8.5 GHz flux densities above and below 500~mJy. Figure~\ref{fig:mi_SI_weakStrong} shows that the relationship between modulation index and structure index holds independently for the two sub-samples, although it is more pronounced for weak sources. We note that there are only 24 weak sources in our sample, and we therefore refrain from repeating the above analysis separately for weak and strong samples.

The results presented above clearly show that persistent, high modulation index ISS is an excellent indicator of compact structure. Such scintillators have highly repeatable astrometric positions, and little structure in their VLBI images.  Clearly those quasars from the MASIV sample which are persistent scintillators and not regularly included in IVS geodetic observations are likely to be good geodetic/astrometric sources.  In Appendix~\ref{app:persistentScints} we list all the MASIV persistent scintillators, including the number of times they have been included in IVS observations.  Most have not been regularly observed. 

\section{Conclusions}
\label{sec:conclusions}

We have examined the relationship between astrometric stability and astrophysical properties of flat-spectrum radio-loud quasars. Astrometric stability was evaluated by fitting confidence ellipses to source positions determined from geodetic VLBI observations using the OCCAM geodetic analysis package. Two astrophysical measures of compactness were used: structure index, which uses VLBI images to quantify how point-like a given source is; and presence of intraday variability in quasar light curves due to interstellar scintillation.

We found that persistent scintillators (i.e. sources that scintillate at all observed epochs) have significantly better position stability than episodic scintillators. The episodic scintillators are in turn more stable than non-scintillators. We also found a strong correlation between scintillation and compactness on VLBI scales, as given by source structure indices. These findings are consistent with the results of \citet{OjhaEA04} who found that scintillating quasars appear much more compact in VLBI images than their non-scintillating counterparts, confirming that it is the compact core which scintillates.

The modulation index of the ISS provides a further proxy for compactness, with the high modulation index scintillators exhibiting least structure. The timescale of the variability is also important: the scintillators with the longest variability timescales are the most compact. To date, scintillation surveys have only been able to identify sources that vary on timescales less than a few days \citep{LovellEA08,KedzioraChudczerEA01}. Our findings suggest that dedicated, high-cadence monitoring campaigns are required to identify the most compact, slow scintillators \citep[e.g.][]{McCullochEA05,TurnerEA12}. 

In summary, we suggest that persistent, high amplitude scintillators are excellent candidates for any future realisation of the celestial reference frame, and should be targeted by future astrometric campaigns. A list of candidate sources is given in Appendix~\ref{app:persistentScints}.

\bibliography{apj-jour,schaapEA13_130429}

\appendix

\section{Persistent scintillators}
\label{app:persistentScints}

In Table~\ref{tab:persistentScints} we list persistent MASIV scintillators (i.e. sources that varied at all four epochs) together with their position stability major axis lengths. We recommend those persistent scintillators which are not at at present included in the current realisation of the celestial reference frame be scheduled for research and development sessions with a view of being included in any future realisations of the ICRF.

\begin{table*}
\caption{MASIV persistent scintillators. Sources without a $\theta_{\rm major}$ value have less than eight position measurements. Superscripts in column 2 indicate whether the source is an: (v) ICRF2 VLBA Calibrator Survey (VCS) source, typically observed only once or twice during IVS research and development sessions; (r) frequently observed non-VCS ICRF2 source, scheduled on a regular basis with the IVS; (n) not an ICRF2 source, and therefore not observed by the IVS.}
\begin{center}
\begin{tabular}{|c|c|c|c|c|c|c|c|c|c|c|c|c|}
\hline
MASIV	&	IERS & RA && & Dec && & $S_{\rm 8.5}$ 	& average &  maximum	& $\theta_{\rm major}$ & 	No. position \\
name	&	name & hr & min & sec & $\circ$ & ' & '' & (mJy)	& mod. index &  mod. index	& ($\mu$as) & 	observations  \\
\hline
J0102+5824	&	0059+581	$^{\rm	r	}$ &	01	&	02	&	45.76	&	+58	&	24	&	11.1	&	1399	&	0.052	&	0.080	&	9.1	&	1764	\\
J0136+4751	&	0133+476	$^{\rm	r	}$ &	01	&	36	&	58.59	&	+47	&	51	&	29.1	&	1699	&	0.023	&	0.042	&	9.1	&	1237	\\
J0150+2646	&	-	$^{\rm	n	}$ &	01	&	50	&	02.80	&	+26	&	46	&	28.0	&	82	&	0.058	&	0.077	&	-	&	0	\\
J0237+2046	&	-	$^{\rm	n	}$ &	02	&	37	&	50.62	&	+20	&	46	&	18.4	&	107.9	&	0.035	&	0.052	&	-	&	0	\\
J0238+1636	&	0235+164	$^{\rm	r	}$ &	02	&	38	&	38.93	&	+16	&	36	&	59.2	&	5340	&	0.022	&	0.030	&	14.4	&	790	\\
J0253+3217	&	0250+320	$^{\rm	v	}$ &	02	&	53	&	33.65	&	+32	&	17	&	20.8	&	112.1	&	0.065	&	0.125	&	-	&	1	\\
J0313+0228	&	0310+022	$^{\rm	v	}$ &	03	&	13	&	13.40	&	+02	&	28	&	35.2	&	127	&	0.057	&	0.069	&	-	&	1	\\
J0328+2552	&	0325+256	$^{\rm	v	}$ &	03	&	28	&	44.34	&	+25	&	52	&	08.4	&	120	&	0.030	&	0.050	&	-	&	0	\\
J0342+3859	&	-	$^{\rm	n	}$ &	03	&	42	&	16.26	&	+38	&	59	&	06.2	&	128.4	&	0.035	&	0.045	&	-	&	0	\\
J0343+3622	&	0340+362	$^{\rm	r	}$ &	03	&	43	&	28.95	&	+36	&	22	&	12.4	&	620	&	0.048	&	0.083	&	-	&	6	\\
J0409+1217	&	0406+121	$^{\rm	r	}$ &	04	&	09	&	22.00	&	+12	&	17	&	39.8	&	657	&	0.029	&	0.044	&	41.6	&	31	\\
J0411+0843	&	-	$^{\rm	n	}$ &	04	&	11	&	33.85	&	+08	&	43	&	11.4	&	116	&	0.048	&	0.054	&	-	&	0	\\
J0419+3955	&	0415+398	$^{\rm	r	}$ &	04	&	19	&	22.55	&	+39	&	55	&	28.9	&	640	&	0.030	&	0.042	&	11.5	&	11	\\
J0453+0128	&	0450+013	$^{\rm	v	}$ &	04	&	53	&	02.23	&	+01	&	28	&	35.6	&	111	&	0.070	&	0.133	&	-	&	1	\\
J0502+1338	&	0459+135	$^{\rm	r	}$ &	05	&	02	&	33.21	&	+13	&	38	&	10.9	&	928	&	0.049	&	0.057	&	-	&	5	\\
J0509+0541	&	0506+056	$^{\rm	v	}$ &	05	&	09	&	25.96	&	+05	&	41	&	35.3	&	684	&	0.050	&	0.069	&	-	&	1	\\
J0605+4030	&	0602+405	$^{\rm	r	}$ &	06	&	05	&	50.85	&	+40	&	30	&	08.0	&	780.8	&	0.014	&	0.018	&	-	&	5	\\
J0643+1238	&	-	$^{\rm	n	}$ &	06	&	43	&	59.85	&	+12	&	38	&	18.0	&	105	&	0.043	&	0.061	&	-	&	0	\\
J0757+0956	&	0754+100	$^{\rm	r	}$ &	07	&	57	&	06.64	&	+09	&	56	&	34.8	&	1377	&	0.034	&	0.058	&	19.0	&	23	\\
J0758+0827	&	-	$^{\rm	n	}$ &	07	&	58	&	28.04	&	+08	&	27	&	09.0	&	119.7	&	0.050	&	0.085	&	-	&	0	\\
J0804+1012	&	-	$^{\rm	n	}$ &	08	&	04	&	07.58	&	+10	&	12	&	13.2	&	129.6	&	0.040	&	0.050	&	-	&	0	\\
J0829+4018	&	0826+404	$^{\rm	n	}$ &	08	&	29	&	35.57	&	+40	&	18	&	59.2	&	123	&	0.124	&	0.175	&	-	&	0	\\
J0854+8034	&	0847+807	$^{\rm	v	}$ &	08	&	54	&	48.58	&	+80	&	34	&	22.3	&	122	&	0.020	&	0.025	&	-	&	1	\\
J0856+7146	&	0851+719	$^{\rm	v	}$ &	08	&	56	&	54.86	&	+71	&	46	&	23.8	&	118	&	0.037	&	0.051	&	-	&	1	\\
J0914+0245	&	0912+029	$^{\rm	r	}$ &	09	&	14	&	37.91	&	+02	&	45	&	59.2	&	693	&	0.023	&	0.052	&	23.0	&	31	\\
J0916+0242	&	-	$^{\rm	n	}$ &	09	&	16	&	41.77	&	+02	&	42	&	52.9	&	106.2	&	0.038	&	0.057	&	-	&	0	\\
J0929+5013	&	0925+504	$^{\rm	r	}$ &	09	&	29	&	15.44	&	+50	&	13	&	35.9	&	692	&	0.044	&	0.062	&	-	&	2	\\
J0946+5020	&	0942+505	$^{\rm	v	}$ &	09	&	46	&	16.04	&	+50	&	20	&	09.3	&	112	&	0.052	&	0.081	&	-	&	1	\\
J0958+4725	&	0955+476	$^{\rm	r	}$ &	09	&	58	&	19.67	&	+47	&	25	&	07.8	&	881	&	0.016	&	0.026	&	14.3	&	1873	\\
J1008+0621	&	1005+066	$^{\rm	v	}$ &	10	&	08	&	00.81	&	+06	&	21	&	21.2	&	674	&	0.024	&	0.026	&	-	&	1	\\
J1024+2332	&	1022+237	$^{\rm	v	}$ &	10	&	24	&	53.63	&	+23	&	32	&	33.9	&	128.5	&	0.041	&	0.057	&	-	&	1	\\
J1049+1429	&	1047+147	$^{\rm	r	}$ &	10	&	49	&	46.32	&	+14	&	29	&	38.5	&	115	&	0.044	&	0.059	&	-	&	0	\\
J1159+2914	&	1156+295	$^{\rm	r	}$ &	11	&	59	&	31.83	&	+29	&	14	&	43.8	&	1205	&	0.046	&	0.058	&	13.6	&	1187	\\
J1247+7046	&	1245+710	$^{\rm	v	}$ &	12	&	47	&	07.55	&	+70	&	46	&	45.1	&	115.6	&	0.027	&	0.033	&	-	&	0	\\
J1328+6221	&	-	$^{\rm	n	}$ &	13	&	28	&	40.56	&	+62	&	21	&	37.0	&	107.1	&	0.061	&	0.080	&	-	&	0	\\
J1417+3818	&	1415+385	$^{\rm	n	}$ &	14	&	17	&	40.44	&	+38	&	18	&	21.1	&	119	&	0.026	&	0.027	&	-	&	0	\\
J1442+0625	&	-	$^{\rm	n	}$ &	14	&	42	&	12.23	&	+06	&	25	&	26.1	&	117.4	&	0.034	&	0.044	&	-	&	0	\\
J1444+0257	&	-	$^{\rm	n	}$ &	14	&	44	&	31.76	&	+02	&	57	&	53.4	&	118.7	&	0.029	&	0.032	&	-	&	0	\\
J1610+7809	&	1613+782	$^{\rm	n	}$ &	16	&	10	&	50.62	&	+78	&	09	&	00.5	&	126	&	0.033	&	0.052	&	-	&	0	\\
J1648+2141	&	-	$^{\rm	n	}$ &	16	&	48	&	17.06	&	+21	&	41	&	05.8	&	105.8	&	0.056	&	0.069	&	-	&	0	\\
J1739+4737	&	1738+476	$^{\rm	r	}$ &	17	&	39	&	57.12	&	+47	&	37	&	58.3	&	829	&	0.015	&	0.019	&	28.0	&	25	\\
J1740+5211	&	1739+522	$^{\rm	r	}$ &	17	&	40	&	36.97	&	+52	&	11	&	43.4	&	1300	&	0.017	&	0.019	&	10.8	&	2001	\\
J1747+4658	&	1746+470	$^{\rm	r	}$ &	17	&	47	&	26.64	&	+46	&	58	&	50.9	&	871	&	0.032	&	0.038	&	9.1	&	19	\\
J1812+5603	&	1812+560	$^{\rm	v	}$ &	18	&	12	&	57.66	&	+56	&	03	&	49.1	&	108	&	0.019	&	0.022	&	-	&	2	\\
J1819+3845	&	1817+387	$^{\rm	v	}$ &	18	&	19	&	26.54	&	+38	&	45	&	01.7	&	128.6	&	0.335	&	0.371	&	-	&	1	\\
J1931+4743	&	-	$^{\rm	n	}$ &	19	&	31	&	16.55	&	+47	&	43	&	41.2	&	122.8	&	0.029	&	0.045	&	-	&	0	\\
J2006+6424	&	2005+642	$^{\rm	r	}$ &	20	&	06	&	17.69	&	+64	&	24	&	45.4	&	973	&	0.018	&	0.029	&	-	&	7	\\
J2016+1632	&	2013+163	$^{\rm	r	}$ &	20	&	16	&	13.86	&	+16	&	32	&	34.1	&	613	&	0.027	&	0.041	&	12.5	&	13	\\
J2113+1121	&	-	$^{\rm	n	}$ &	21	&	13	&	54.72	&	+11	&	21	&	25.4	&	127.1	&	0.047	&	0.073	&	-	&	0	\\
J2203+1725	&	2201+171	$^{\rm	r	}$ &	22	&	03	&	26.89	&	+17	&	25	&	48.2	&	990	&	0.028	&	0.034	&	-	&	5	\\
J2212+2759	&	2210+277	$^{\rm	v	}$ &	22	&	12	&	39.10	&	+27	&	59	&	38.4	&	115	&	0.029	&	0.039	&	-	&	0	\\
J2237+4216	&	2234+420	$^{\rm	v	}$ &	22	&	37	&	04.20	&	+42	&	16	&	48.2	&	122.8	&	0.098	&	0.120	&	-	&	0	\\
J2241+4120	&	2238+410	$^{\rm	v	}$ &	22	&	41	&	07.20	&	+41	&	20	&	11.6	&	825.6	&	0.036	&	0.057	&	-	&	1	\\
J2303+1431	&	2300+142	$^{\rm	v	}$ &	23	&	03	&	09.95	&	+14	&	31	&	41.3	&	121.9	&	0.035	&	0.048	&	-	&	0	\\
J2311+4543	&	2309+454	$^{\rm	r	}$ &	23	&	11	&	47.41	&	+45	&	43	&	56.0	&	610	&	0.018	&	0.022	&	13.2	&	12	\\
J2325+3957	&	2322+396	$^{\rm	n	}$ &	23	&	25	&	17.87	&	+39	&	57	&	36.5	&	115.5	&	0.040	&	0.046	&	-	&	0	\\
\hline
\end{tabular}
\end{center}
\label{tab:persistentScints}
\end{table*}

%
%

\section*{Acknowledgments}

R.S. is grateful to the University of Tasmania for a Dean's Summer Research Scholarship. S.S. thanks the Australian Research Council for a Super Science Fellowship. We are grateful to Dave Jauncey and Cliff Senkbeil for discussions of an earlier version of this analysis. This research has made use of material from the Bordeaux VLBI Image Database (BVID).

\bsp

\label{lastpage}

\end{document}